\documentclass[aps,prd,preprint,superscriptaddress,showpacs,floatfix]{revtex4}

\usepackage[dvips]{graphics}
\usepackage[dvips]{graphicx}
\usepackage{psfrag}
\usepackage{longtable}
\usepackage{amsfonts}

\begin{document}


\preprint{BNL-HET-02/30,COLO-HEP-484}
\preprint{December 2002}

\title{Symmetric Textures in SO(10) and LMA Solution for Solar Neutrinos} 


\author{Mu-Chun Chen}
\email[]{chen@quark.phy.bnl.gov}
\affiliation{High Energy Theory Group, Department of Physics, 
Brookhaven National Laboratory, Upton, NY 11973, U.S.A.}
\author{K.T. Mahanthappa}
\email[]{ktm@verb.colorado.edu}
\affiliation{Department of Physics, University of Colorado, 
Boulder, CO80309-0390, U.S.A.}



\begin{abstract}
We analyze a model based on SUSY $SO(10)$ combined with $SU(2)$
family symmetry and symmetric mass matrices constructed by the authors recently. 
Previously, only the parameter space for the LOW and vacuum oscillation (VO) 
solutions was investigated. We indicate in this note   
the parameter space which leads to large mixing angle (LMA) 
solution to the solar neutrino problem with a slightly 
modified effective neutrino mass matrix. The symmetric mass textures 
arising from the left-right symmetry breaking and the $SU(2)$ symmetry breaking 
give rise to very good predictions for the quark and lepton masses and mixing 
angles. The prediction of our model for the $|U_{e\nu_{3}}|$ element in 
the Maki-Nakagawa-Sakata (MNS) matrix is close to  
the sensitivity of current experiments; thus the validity of our model can be 
tested in the near future. We also investigate the correlation between 
the $|U_{e\nu_{3}}|$ element and $\tan^{2}\theta_{\odot}$ in a general 
two-zero neutrino mass texture. 
\end{abstract}



\maketitle


The recently reported measurements from KamLAND reactor 
experiment~\cite{unknown:2002dm} confirmed the large mixing angle 
(LMA) solution to be the unique oscillation solution to the solar neutrino problem 
at $4.7 \; \sigma$ level~\cite{Maltoni:2002aw,Bahcall:2002ij,Fogli:2002au}.
The global analysis including Solar + KamLAND + CHOOZ 
indicate the following allowed region at 
$3\sigma$~\cite{Maltoni:2002aw},
\begin{eqnarray}
5.1 \times 10^{-5} < \Delta m_{21}^2 < 9.7 \times 10^{-5} eV^{2}
\\
0.29 \le \tan^{2}\theta_{12} \le 0.86
\\
(0.70 \le \sin^{2} 2\theta_{12} \le 0.994)
\end{eqnarray}
The allowed regions at $3\sigma$ level based on a global fit including 
SK + Solar + CHOOZ for the atmospheric parameters and the CHOOZ 
angle are~\cite{Gonzalez-Garcia:2002dz}
\begin{eqnarray}
1.4 \times 10^{-3} < \Delta m_{32}^2 < 6.0 \times 10^{-3} eV^{2}
\\
0.4 \le tan^{2} \theta_{23} \le 3.0\\
(0.82 < \sin^{2} 2\theta_{23})
\\
\sin^{2} \theta_{13} < 0.06
\end{eqnarray}
There have been a few $SO(10)$ models constructed aiming to accommodate the 
observed neutrino masses and 
mixing angles (see for example, 
\cite{Chen:2000fp,Chen:2001pr,Blazek:1999ue}). 
By far, the LMA solution is the most difficult to obtain. 
Most of the models in the literature assume the mass matrices ``lopsided''. 
In our model based on SUSY $SO(10) \times SU(2)$~\cite{Chen:2000fp,Chen:2001pr} 
(referred to ``CM'' herein), we consider {\it symmetric} mass matrices which are 
resulting from the left-right symmetric breaking of $SO(10)$ and the breaking of 
family symmetry $SU(2)$. Previously, we studied the parameter space for the 
LOW and VO solutions to the solar neutrino problem in our model. 
In view of the KamLAND result, we re-analyze our model and find the 
parameter space for the LMA solution.

The details of our model based on $SO(10) \times SU(2)_{F}$ are contained 
in CM. The following is an outline of its salient features. In order to
specify the superpotential uniquely, we invoke 
$Z_{2} \times Z_{2} \times Z_{2}$ discrete symmetry. The matter fields are
\begin{displaymath} 
\psi_{a} \sim (16,2)^{-++} \quad (a=1,2), \qquad 
\psi_{3} \sim (16,1)^{+++} 
\end{displaymath}
where $a=1,2$ and the subscripts refer to family indices; the superscripts 
$+/-$ refer to $(Z_{2})^{3}$ charges. The Higgs fields which break $SO(10)$
and give rise to mass matrices upon acquiring VEV's are
\begin{eqnarray}
(10,1):\quad & T_{1}^{+++}, \quad T_{2}^{-+-},\quad
T_{3}^{--+}, \quad T_{4}^{---}, \quad T_{5}^{+--} \nonumber\\ 
(\overline{126},1):\quad & \overline{C}^{---}, \quad \overline{C}_{1}^{+++},
\quad \overline{C}_{2}^{++-}  
\end{eqnarray}
Higgs representations $10$ and $\overline{126}$ give rise to Yukawa couplings
to the matter fields which are symmetric under the interchange of family
indices. $SO(10)$ is broken through the left-right symmetry breaking chain
\begin{equation}
\label{eq:SB}
\begin{array}{lll}
SO(10) & \longrightarrow & SU(4) \times SU(2)_{L} \times SU(2)_{R}\\
 & \longrightarrow & SU(3) \times SU(2)_{L} \times SU(2)_{R} \times
U(1)_{B-L}\\  
& \longrightarrow & SU(3) \times SU(2)_{L} \times U(1)_{Y} \\
& \longrightarrow & SU(3) \times U(1)_{EM} 
\end{array}
\end{equation}
The $SU(2)$ family symmetry~\cite{Barbieri:1997ww} is broken in two steps and
the mass hierarchy is produced using the Froggatt-Nielsen 
mechanism:
\begin{equation}
\label{eq:steps} 
SU(2) \stackrel{\epsilon M}{\longrightarrow} 
U(1) \stackrel{\epsilon' M}{\longrightarrow}
nothing
\end{equation}
where $M$ is the UV-cutoff of the effective theory above which the family
symmetry is exact, and $\epsilon M$ and $\epsilon^{'} M$ are the VEV's
accompanying the flavon fields given by
\begin{eqnarray}
(1,2): \quad & \phi_{(1)}^{++-}, \quad \phi_{(2)}^{+-+}, \quad \Phi^{-+-}
\nonumber\\ 
(1,3): \quad & S_{(1)}^{+--}, \quad S_{(2)}^{---}, \quad
\Sigma^{++-} 
\end{eqnarray}
The various aspects of VEV's of Higgs and flavon fields are given in CM.

The superpotential of our model is
\begin{equation}
W = W_{Dirac} + W_{\nu_{RR}}
\end{equation}
\begin{eqnarray}
W_{Dirac}=\psi_{3}\psi_{3} T_{1}
 + \frac{1}{M} \psi_{3} \psi_{a}
\left(T_{2}\phi_{(1)}+T_{3}\phi_{(2)}\right)
\nonumber\\
+ \frac{1}{M} \psi_{a} \psi_{b} \left(T_{4} + \overline{C}\right) S_{(2)}
+ \frac{1}{M} \psi_{a} \psi_{b} T_{5} S_{(1)}
\nonumber\\
W_{\nu_{RR}}=\psi_{3} \psi_{3} \overline{C}_{1} 
+ \frac{1}{M} \psi_{3} \psi_{a} \Phi \overline{C}_{2}
+ \frac{1}{M} \psi_{a} \psi_{b} \Sigma \overline{C}_{2}
\end{eqnarray}
The mass matrices then can be read from the superpotential to be
\begin{eqnarray}
M_{u,\nu_{LR}} & = &
\left( \begin{array}{ccc}
0 & 0 & \left<10_{2}^{+} \right> \epsilon'\\
0 & \left<10_{4}^{+} \right> \epsilon & \left<10_{3}^{+} \right> \epsilon \\
\left<10_{2}^{+} \right> \epsilon' & \left<10_{3}^{+} \right> \epsilon &
\left<10_{1}^{+} \right>
\end{array} \right)
\nonumber\\
 & = & 
\left( \begin{array}{ccc}
0 & 0 & r_{2} \epsilon'\\
0 & r_{4} \epsilon & \epsilon \\
r_{2} \epsilon' & \epsilon & 1
\end{array} \right) M_{U}
\end{eqnarray}
\begin{eqnarray}
M_{d,e} & = & 
\left(\begin{array}{ccc}
0 & \left<10_{5}^{-} \right> \epsilon' & 0 \\
\left<10_{5}^{-} \right> \epsilon' &  (1,-3)\left<\overline{126}^{-} \right>
\epsilon & 0\\ 0 & 0 & \left<10_{1}^{-} \right>
\end{array} \right)
\nonumber\\
 & = & 
\left(\begin{array}{ccc}
0 & \epsilon' & 0 \\
\epsilon' &  (1,-3) p \epsilon & 0\\
0 & 0 & 1
\end{array} \right) M_{D}
\end{eqnarray}
where
$M_{U} \equiv \left<10_{1}^{+} \right>$, 
$M_{D} \equiv \left<10_{1}^{-} \right>$, 
$r_{2} \equiv \left<10_{2}^{+} \right> / \left<10_{1}^{+} \right>$, 
$r_{4} \equiv \left<10_{4}^{+} \right> / \left<10_{1}^{+} \right>$ and
$p \equiv \left<\overline{126}^{-}\right> / \left<10_{1}^{-} \right>$.
The right-handed neutrino mass matrix is  
\begin{eqnarray}
M_{\nu_{RR}} & = &  
\left( \begin{array}{ccc}
0 & 0 & \left<\overline{126}_{2}^{'0} \right> \delta_{1}\\
0 & \left<\overline{126}_{2}^{'0} \right> \delta_{2} 
& \left<\overline{126}_{2}^{'0} \right> \delta_{3} \\ 
\left<\overline{126}_{2}^{'0} \right> \delta_{1}
& \left<\overline{126}_{2}^{'0} \right> \delta_{3} &
\left<\overline{126}_{1}^{'0} \right> \end{array} \right)
\nonumber\\
 & = & 
\left( \begin{array}{ccc}
0 & 0 & \delta_{1}\\
0 & \delta_{2} & \delta_{3} \\ 
\delta_{1} & \delta_{3} & 1
\end{array} \right) M_{R}
\label{Mrr}
\end{eqnarray}
with $M_{R} \equiv \left<\overline{126}^{'0}_{1}\right>$.
Here the superscripts $+/-/0$ refer to the sign of the hypercharge. 
It is to be noted that there is a factor of $-3$ difference between the $(22)$
elements of mass matrices $M_{d}$ and $M_{e}$. This is due to the CG
coefficients associated with $\overline{126}$; as a consequence, we obtain the
phenomenologically viable Georgi-Jarlskog relation.
We then parameterize the Yukawa matrices as follows, after removing 
all the non-physical phases by rephasing various matter fields: 
\begin{eqnarray}
Y_{u, \nu_{LR}}=\left(
\begin{array}{ccc}
0 & 0 & a\\
0 & b e^{i\theta} & c\\
a & c & 1
\end{array}
\right) d
\\
Y_{d,e}=\left(
\begin{array}{ccc}
0 & e e^{-i\xi} & 0 \\
e e^{i\xi} & (1,-3) f & 0 \\
0 & 0 & 1
\end{array}
\right) h
\label{phaseremoved}
\end{eqnarray}
This is one of the five sets of symmetric texture combinations (labeled set
(v)) proposed by Ramond, Roberts and Ross~\cite{Ramond:1993kv}. 

\begin{figure}
\psfrag{U_13}[][]{$|U_{e\nu_{3}}|^{2}$}
\psfrag{tan}[][]{$\tan^{2}\theta_{\odot}$}
\psfrag{t}[][]{$\scriptstyle n=1 \qquad \qquad$}
\psfrag{t32}[][]{$\scriptstyle n=3/2 \quad \qquad$}
\psfrag{t2}[][]{$\scriptstyle n=2 \qquad \qquad$}
\psfrag{mh}[][]{$\scriptscriptstyle \Delta m_{\odot} 
= 1 \times 10^{-4} eV^{2} \qquad \qquad \qquad \;$}
\psfrag{mm}[][]{$\scriptscriptstyle \Delta m_{\odot} 
= 7.3 \times 10^{-5} eV^{2} \qquad \qquad \qquad$}
\psfrag{ml}[][]{$\scriptscriptstyle \Delta m_{\odot} 
= 5 \times 10^{-5} eV^{2} \qquad \qquad \qquad \;$}
{\center
\includegraphics[scale=0.65, angle=270]{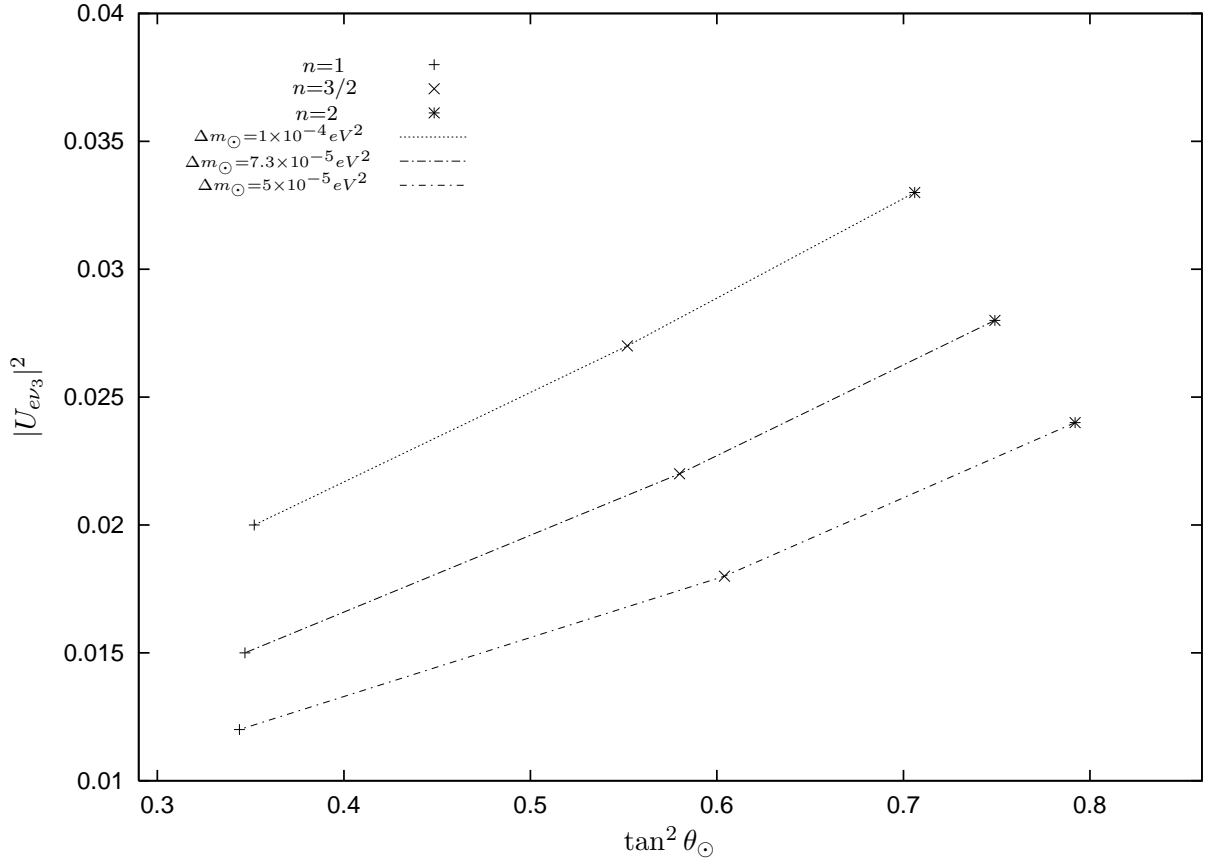}}
\caption{\label{corelate}Correlation between $|U_{e\nu_{3}}|^{2}$ and 
$\tan^{2}\theta_{\odot}$ for different values of $n$. 
The value of $\Delta m_{atm}^{2}$ is $2.8 \times 10^{-3} eV^{3}$. 
The dotted line corresponds to the upper bound 
$\Delta m_{\odot}^{2} = 10^{-4} eV^{2}$; the dotted-long-dashed 
line corresponds to the best fit value $\Delta m_{\odot}^{2} 
= 7.3 \times 10^{-5} eV^{2}$; the dotted-short-dashed 
line corresponds to the lower bound $\Delta m_{\odot}^{2} 
= 5 \times 10^{-5} eV^{2}$. So a generic viable prediction of the texture 
given in Eq.~[\ref{texture}] is in the region bounded by the dotted line and 
the dotted-short-dashed line.}
\end{figure}

We use the following inputs at 
$M_{Z}=91.187 \; GeV$~\cite{Fusaoka:1998vc,Hocker:2001xe}:
\begin{eqnarray}
m_{u} & = & 2.32 \; MeV (2.33^{+0.42}_{-0.45})\nonumber\\ 
m_{c} & = & 677 \; MeV (677^{+56}_{-61})\nonumber\\
m_{t} & = & 182 \; GeV (181^{+}_{-}13) \nonumber\\
m_{e} & = & 0.485 \; MeV (0.486847)\nonumber\\
m_{\mu} & = & 103 \; MeV (102.75)\nonumber\\
m_{\tau} & = & 1.744 \; GeV (1.7467) \nonumber\\
\vert V_{us} \vert & = & 0.222 (0.219-0.224)\nonumber\\
\vert V_{ub} \vert & = & 0.0039 (0.002-0.005)\nonumber\\
\vert V_{cb} \vert & = & 0.036 (0.036-0.046)\nonumber
\end{eqnarray} 
where the values extrapolated from experimental data are given inside the
parentheses. 
These values correspond to the following set of input parameters
at the GUT scale,  $M_{GUT} = 1.03 \times 10^{16} \; GeV$: 
\begin{eqnarray} 
& a = 0.00246, \quad b =  3.50 \times 10^{-3}\nonumber\\
& c = 0.0320, \quad  d =  0.650\nonumber\\
& \theta  = 0.110 \nonumber\\
& e  = 4.03 \times 10^{-3}, \quad f  =  0.0195 \nonumber\\
& h =  0.0686, \quad \xi  =  -0.720 \nonumber\\
& g_{1} =  g_{2} = g_{3}  =  0.746
\end{eqnarray}
the one-loop renormalization group equations for the MSSM spectrum with three
right-handed neutrinos 
are solved numerically down to the effective
right-handed neutrino mass scale, $M_{R}$. At $M_{R}$, the seesaw mechanism  
is implemented. With the constraints $|m_{\nu_{3}}| \gg |m_{\nu_{2}}|, \; |m_{\nu_{1}}|$ and 
maximal mixing in the atmospheric sector, the up-type mass texture leads us 
to choose the following effective neutrino mass matrix
\begin{equation}\label{mll}
M_{\nu_{LL}} = \left(
\begin{array}{ccc}
0 & 0 & t\\
0 & 1 & 1+t^{3/2}\\
t & 1+t^{3/2} & 1
\end{array}
\right)\frac{d^{2}v_{u}^{2}}{M_{R}}
\end{equation}
and from the seesaw formula we obtain
\begin{eqnarray}
\label{delta}
\delta_{1} & = & \frac{a^{2}}{c^{2}t+a^{2}(2t^{1/2}+t^{2})+2a(1-c(1+t^{3/2}))}
\nonumber\\
\delta_{2} & = &
\frac{b^{2}t e^{2i\theta}}{c^{2}t+a^{2}(2t^{1/2}+t^{2})+2a(1-c(1+t^{3/2}))}
\nonumber\\
\delta_{3} & = &
\frac{-a(be^{i\theta}(1+t^{3/2})-c)+bct e^{i\theta}}
{c^{2}t+a^{2}(2t^{1/2}+t^{2})+2a(1-c(1+t^{3/2}))}
\end{eqnarray}
We then solve the two-loop RGE's for the MSSM spectrum 
down to the SUSY breaking scale, taken to be $m_{t}(m_{t})=176.4 \; GeV$, and
then the SM RGE's from $m_{t}(m_{t})$ to the weak scale, $M_{Z}$. 
We assume that 
$\tan \beta \equiv v_{u}/v_{d} = 10$,  with 
$v_{u}^{2} + v_{d}^{2} = (246/\sqrt{2} \; GeV) ^{2}$. At the weak scale
$M_{Z}$, the predictions for 
$\alpha_{i}
\equiv g_{i}^{2}/4\pi$ are   
\begin{displaymath} 
\alpha_{1}=0.01663,
\quad \alpha_{2}=0.03374, 
\quad \alpha_{3}=0.1242 
\end{displaymath}
These values compare very well with the values extrapolated to $M_{Z}$ from the
experimental data, 
$(\alpha_{1},\alpha_{2},\alpha_{3})=
(0.01696,0.03371,0.1214 \pm 0.0031)$.
The predictions at the weak scale $M_{Z}$ for the
charged fermion masses, CKM matrix elements and strengths of CP violation, 
are summarized in Table. I of Ref.~\cite{Chen:2001pr}. 
Using the mass square difference in the atmospheric sector 
$\Delta m_{atm}^{2}=2.78 \times 10^{-3} \; eV^{2}$ and 
the mass square difference for the LMA solution 
$\Delta m_{\odot}^{2}=7.25 \times 10^{-5} \; eV^{2}$ as input 
parameters, we determine 
$t = 0.35$ and $M_{R} = 5.94 \times 10^{12} GeV$, and correspondingly  
$(\delta_{1},\delta_{2},\delta_{3}) 
= (0.00119,0.000841 e^{i \; (0.220)},0.0211 e^{-i \;(0.029)})$. We obtain 
the following predictions in the neutrino sector: 
The three mass eigenvalues are give by  
\begin{equation}
(m_{\nu_{1}},m_{\nu_{2}},m_{\nu_{3}}) = (0.00363,0.00926,0.0535) \; eV
\end{equation}
The prediction for the MNS matrix is
\begin{equation}
\vert U_{MNS} \vert = 
\left(
\begin{array}{ccc}
0.787 & 0.599 & 0.149\\
0.508 & 0.496 & 0.705\\
0.350 & 0.629 & 0.694
\end{array}
\right)
\end{equation}
which translates into the mixing angles in the atmospheric, 
solar and reactor sectors,
\begin{eqnarray}
\sin^{2} 2 \theta_{atm} & \equiv & \frac{
4 \vert U_{\mu \nu_{3}} \vert^{2} |U_{\tau \nu_{3}}|^{2}}
{(1-|U_{e\nu_{3}}|^2)^{2}}
= 1
\nonumber\\
\tan^{2} \theta_{atm} & \equiv & \frac{\vert U_{\mu \nu_{3}}\vert^{2}}
{|U_{\tau \nu_{3}}|^{2}} = 1.03
\nonumber\\
\sin^{2} 2 \theta_{\odot} & \equiv & \frac{
4 \vert U_{e \nu_{1}} \vert^{2} \vert U_{e \nu_{2}} \vert^{2}}
{(1-|U_{e\nu_{3}}|^2)^{2}} = 0.93
\nonumber\\
\tan^{2} \theta_{\odot} & \equiv & \frac{\vert U_{e \nu_{2}}\vert^{2}}
{|U_{e \nu_{1}}|^{2}} = 0.58
\nonumber\\
\sin^{2}\theta_{13} & = & |U_{e\nu_{3}}|^{2} = 0.022
\end{eqnarray}
To our precision, the atmospheric mixing angle is maximal, 
while the solar angle is within the allowed region at $1 \sigma$ level 
($0.37 \le \tan^{2}\theta_{\odot} \le 0.60$~\cite{Bahcall:2002ij}). 
We comment that $M_{\nu_{LL}}$ given in Eq.~[\ref{mll}] is a special case 
of a two-zero texture
\begin{equation}
\left(\begin{array}{ccc}
0 & 0 & \ast\\
0 & \ast & \ast\\
\ast & \ast & \ast
\end{array}
\right)
\end{equation}
first proposed in~\cite{Chen:2000fp} in which the elements in $(23)$ block 
are taken to have equal strengths to accommodate near bi-maximal mixing. 
Here we consider a slightly different case 
\begin{equation}\label{texture}
\left(\begin{array}{ccc}
0 & 0 & t\\
0 & 1 & 1+t^{n}\\
t & 1+t^{n} & 1
\end{array}
\right)
\end{equation}
This modification is needed in order to accommodate a large, but non-maximal 
solar angle in the so-called ``light side'' region 
($0 < \theta < \pi/4$)~\cite{deGouvea:2000cq}.  
We find that it is possible to obtain the LMA solution at $3\sigma$ level   
with $n$ ranging from $1$ to $2$. To obtain the LMA solution within 
the allowed region at $1\sigma$ level, 
we have considered above $n=3/2$. The correlation between 
$|U_{e\nu_{3}}|^{2}$ and 
$\tan^{2}\theta_{\odot}$ for different values of $n$ is plotted 
in Fig.~[\ref{corelate}].
The prediction of our model for the strengths of CP violation in 
the lepton sector are
\begin{eqnarray}
J_{CP}^{l} \equiv Im\{ U_{11} U_{12}^{\ast} U_{21}^{\ast} U_{22} \}
= -0.00690
\nonumber\\
(\alpha_{31},\alpha_{21}) = (0.490,-2.29)
\end{eqnarray}
Using the predictions for the neutrino masses, mixing angles and the two Majorana phases, 
$\alpha_{31}$ and $\alpha_{21}$, the matrix element for the neutrinoless double 
$\beta$ decay can be calculated and is given by   
$\vert < m > \vert = 2.22 \times 10^{-3} \; eV$.
Masses of the heavy right-handed neutrinos are
$(M_{1},M_{2},M_{3})=(1.72 \times 10^{7},2.44 \times 10^{9},5.94 \times 10^{12})GeV$. 
As in the case of LOW and VO solutions in our model~\cite{Chen:2001pr}, 
the amount of baryogenesis due to the decay of heavy right-handed neutrinos 
is too small to account for 
the observed amount. Thus another mechanism for baryogenesis is needed in our model. 
The prediction for the $\sin^{2}\theta_{13}$ value is $0.022$, 
in agreement with the current bound $0.06$. Because our prediction for 
$\sin^{2}\theta_{13}$ is very close to the present sensitivity 
of the experiment, the validity of our model can be tested in 
the foreseeable future.

\begin{acknowledgments}
We would like to thank Bob Shrock for pointing out to us that 
the allowed region for $\sin^{2} \theta_{13}$ is larger than 
what we previously thought. We would also like to thank Andre 
de Gouvea for his helpful communication. 
M-CC and KTM are supported, in part, 
by the U.S. Department of Energy under Grant No. DE-AC02-76CH00016 and 
DE-FG03-95ER40892, respectively.
\end{acknowledgments}

\bibliography{cp.commt}

\end{document}